\documentclass[12pt]{article}
\usepackage{epsfig}

\makeatletter
\@addtoreset{equation}{section}

\renewcommand{\thefootnote}{\fnsymbol{footnote}}
\makeatother

\begin{document}
\newcommand{\p}[1]{(\ref{#1})}
\newcommand {\beq}{\begin{eqnarray}}
\newcommand {\eeq}{\end{eqnarray}}
\newcommand {\non}{\nonumber\\}
\newcommand {\eq}[1]{\label {eq.#1}}
\newcommand {\defeq}{\stackrel{\rm def}{=}}
\newcommand {\gto}{\stackrel{g}{\to}}
\newcommand {\hto}{\stackrel{h}{\to}}
\newcommand {\1}[1]{\frac{1}{#1}}
\newcommand {\2}[1]{\frac{i}{#1}}
\newcommand {\thb}{\bar{\theta}}
\newcommand {\ps}{\psi}
\newcommand {\psb}{\bar{\psi}}
\newcommand {\ph}{\varphi}
\newcommand {\phs}[1]{\varphi^{*#1}}
\newcommand {\sig}{\sigma}
\newcommand {\sigb}{\bar{\sigma}}
\newcommand {\Ph}{\Phi}
\newcommand {\Phd}{\Phi^{\dagger}}
\newcommand {\Sig}{\Sigma}
\newcommand {\Phm}{{\mit\Phi}}
\newcommand {\eps}{\varepsilon}
\newcommand {\del}{\partial}
\newcommand {\dagg}{^{\dagger}}
\newcommand {\pri}{^{\prime}}
\newcommand {\prip}{^{\prime\prime}}
\newcommand {\pripp}{^{\prime\prime\prime}}
\newcommand {\prippp}{^{\prime\prime\prime\prime}}
\newcommand {\pripppp}{^{\prime\prime\prime\prime\prime}}
\newcommand {\delb}{\bar{\partial}}
\newcommand {\zb}{\bar{z}}
\newcommand {\mub}{\bar{\mu}}
\newcommand {\nub}{\bar{\nu}}
\newcommand {\lam}{\lambda}
\newcommand {\lamb}{\bar{\lambda}}
\newcommand {\kap}{\kappa}
\newcommand {\kapb}{\bar{\kappa}}
\newcommand {\xib}{\bar{\xi}}
\newcommand {\ep}{\epsilon}
\newcommand {\epb}{\bar{\epsilon}}
\newcommand {\Ga}{\Gamma}
\newcommand {\rhob}{\bar{\rho}}
\newcommand {\etab}{\bar{\eta}}
\newcommand {\chib}{\bar{\chi}}
\newcommand {\tht}{\tilde{\th}}
\newcommand {\zbasis}[1]{\del/\del z^{#1}}
\newcommand {\zbbasis}[1]{\del/\del \bar{z}^{#1}}
\newcommand {\vecv}{\vec{v}^{\, \prime}}
\newcommand {\vecvd}{\vec{v}^{\, \prime \dagger}}
\newcommand {\vecvs}{\vec{v}^{\, \prime *}}
\newcommand {\alpht}{\tilde{\alpha}}
\newcommand {\xipd}{\xi^{\prime\dagger}}
\newcommand {\pris}{^{\prime *}}
\newcommand {\prid}{^{\prime \dagger}}
\newcommand {\Jto}{\stackrel{J}{\to}}
\newcommand {\vprid}{v^{\prime 2}}
\newcommand {\vpriq}{v^{\prime 4}}
\newcommand {\vt}{\tilde{v}}
\newcommand {\vecvt}{\vec{\tilde{v}}}
\newcommand {\vecpht}{\vec{\tilde{\phi}}}
\newcommand {\pht}{\tilde{\phi}}
\newcommand {\goto}{\stackrel{g_0}{\to}}
\newcommand {\tr}{{\rm tr}\,}
\newcommand {\GC}{G^{\bf C}}
\newcommand {\HC}{H^{\bf C}}
\newcommand{\vs}[1]{\vspace{#1 mm}}
\newcommand{\hs}[1]{\hspace{#1 mm}}
\newcommand{\al}{\alpha}
\newcommand{\be}{\beta}
\newcommand{\Lam}{\Lambda}
\newcommand{\kahler}{K\"ahler }
\newcommand{\con}[1]{{\Gamma^{#1}}}
\thispagestyle{empty}
\begin{flushright}
{\tt hep-th/0312037} \\
December, 2003 \\
\end{flushright}
\vspace{2mm}
\begin{center}
{\Large
{\bf Massive Nonlinear Sigma Models and BPS Domain Walls\\
}
\vspace{2mm}
{\bf
in Harmonic Superspace
}
}
\\[8mm]
\vspace{3mm}

\normalsize
  {\large \bf
  Masato~Arai~$^{a}$}
\footnote{\it
arai@fzu.cz}
,
 {\large \bf
Evgeny Ivanov~$^{b}$}
\footnote{\it
eivanov@thsun1.jinr.ru
}
{\large \bf and
Jiri Niederle~$^{a}$}
\footnote{\it
niederle@fzu.cz
}

\vskip 1.5em

{$^{a}$ \it Institute of Physics, AS CR,
  182 21, Praha 8, Czech Republic \\

  $^{b}$Bogoliubov Laboratory of Theoretical Physics, JINR,
  Dubna, \\
  141 980 Moscow region, Russia\\
}
\vspace{3mm}
{\bf Abstract}\\[5mm]
{\parbox{13cm}{\hspace{5mm}
\noindent Four-dimensional massive ${\cal N}=2$ nonlinear sigma models
 and BPS wall solutions are studied
 in the off-shell harmonic superspace approach
 in which ${\cal N} =2$ supersymmetry is manifest.
The general nonlinear sigma model can be described by
 an analytic harmonic potential which is the hyper-K{\"a}hler
 analog of the K{\"a}hler potential
 in ${\cal N}=1$ theory.
We examine the massive nonlinear sigma model with
 multi-center four-dimensional target hyper-K\"ahler metrics and derive
 the corresponding  BPS equation.
We study in some detail two particular cases with the Taub-NUT and
double Taub-NUT metrics. The latter embodies, as its two separate limits,
 both Taub-NUT and Eguchi-Hanson metrics. We find that domain wall
 solutions exist only in the double Taub-NUT case including its 
 Eguchi-Hanson limit.
}}
\end{center}
\vfill
\newpage
\setcounter{page}{1}
\setcounter{footnote}{0}
\renewcommand{\thefootnote}{\arabic{footnote}}
%
%
\section{Introduction}

It is well known that topological solutions are of importance in various
 areas of particle physics.
Recently, there was renewed interest in such solutions because of
their crucial role in the brane world scenario \cite{HW,LED,RS}.
In this scenario, our world is assumed to be realized on topological
 objects like
 domain walls or brane-junctions.
Investigating the quantum fluctuation of the domain wall,
 it was found that zero modes are localized on the wall \cite{rubakov}
 and the low energy theory becomes a theory on the wall.
In other words, domain wall background gives rise to some kind
 of the dimensional reduction as an alternative to the standard Kaluza-Klein
 compactification \cite{klein}.
Supersymmetry (SUSY) can also be implemented
 in these models, and it is actually a powerful device
 for constructing their topological solutions.
In SUSY theories, these often appear as the BPS
 states \cite{BPS} which spontaneously
 break half of the original SUSY \cite{WittenOlive}. Therefore,
 they are called $\frac{1}{2}$ BPS ones \cite{CGR}.
Viewing the four-dimensional world as a domain wall, 
 we are led to deal with SUSY theories in five dimensions.
The minimal possibility is
 ${\cal N}=1, d=5$ SUSY possessing eight supercharges.

SUSY with eight supercharges is very restrictive.
For instance, in theories involving only massless scalar multiplets
 (hypermultiplets), non-trivial interactions can only arise
 from nonlinearities in kinetic terms.
Prior to studying the genuine five-dimensional
 theories with
 hypermultiplets, it is instructive to start with
 similar SUSY theories in four dimensions,
 i.e., ${\cal N}=2, d=4$ theories.
Actually, in ${\cal N}=1,~d=5$ and ${\cal N}=2,~d=4$ theories,
 the hypermultiplets contain the same number of on-shell components,
 viz., two complex scalars and one Dirac fermion.
This analysis of the four-dimensional theory could then be of help 
 in studying
 the brane world scenarios based on
 SUSY theories in higher dimensions \cite{AFNS,EFNS}.

With regard to rigid ${\cal N}=2$ SUSY the target manifold of
 the hypermultiplet $d=4$ sigma models
 must be hyper-K{\"a}hler (HK) \cite{AF1}.
In these theories, the scalar potential can be obtained only if
 the hypermultiplets acquire masses by the Scherk-Schwarz mechanism
 \cite{SS} because of the appearance of central
 charges in the ${\cal N}=2$ Poincar\'e superalgebra \cite{AF2}.
The form of the potential
 is specified by the norm of the Killing vector of the target manifold
 isometry
 whose generator is identified with the central charge \cite{GTT1}
(actually, in Ref. \cite{GTT1}, the massive HK model on an arbitrary toric HK
 manifold was given in the component formalism
 in four-dimensional space-time).
In Ref. \cite{GTT2}, the parallel domain walls solution for the 
 massive $T^*{\bf C}P^{N-1}$
 HK sigma model was found and its moduli space was constructed.
In the same model, dynamics of the domain wall solutions was also studied
 \cite{To} and the calculation of a number of zero modes was 
 performed \cite{Le},
 using the 
 index theorem. In Ref. \cite{ANNS}, a massive nonlinear sigma model with the
 $T^*{\bf C}P^{1}$ target was studied in the ${\cal N}=1$ and
 ${\cal N}=2$ superfield approaches and its domain
 wall solutions were examined.
In \cite{ANS}, 
 the massive $T^*{\bf C}P^{N-1}$ HK model was extended to 
 a massive nonlinear HK
 model with the cotangent bundle over complex Grassmann manifold
 as the target space
 and to some generalization of the latter.
The structure of vacua in these models was examined. 
The lump and Q-lump solutions in the HK nonlinear sigma
 models were also considered in \cite{AT,GPTT,NNS2,PT}.
All models of this type can be called {\it massive} nonlinear HK sigma models.

Some massive nonlinear sigma models were studied in an on-shell
 approach, i.e., by taking account of physical fields
 only \cite{GTT1,GTT2,AT,GPTT,NNS2,PT}.
On the other hand, more appropriate for such a study
is an off-shell formalism. It provides a powerful tool of constructing models with the domain wall
 and brane-junction solutions \cite{NNS1},
 as well as a low-energy effective action
 around the wall \cite{sakamura}.

The most natural description of ${\cal N}=2, d=4$ SUSY field theories
is achieved in the harmonic superspace (HSS) \cite{hss1}.
The HSS approach
is the only one to allow superfield formulations of ${\cal N}=2$ SUSY
theories with all supersymmetries being manifest and off-shell.
In the HSS approach, any HK nonlinear sigma model can be described by
one analytic function
which is the HK analog of K{\"a}hler potential.
This analytic function (HK potential) embodies self-interactions of
hypermultiplets. The massless nonlinear
sigma models in the HSS formalism were studied in \cite{TNhss,EHhss,hss2}.
In particular, in \cite{hss2},
 it was shown that the component action obtained from the
 general massless nonlinear sigma model in HSS coincides with that
 given in \cite{AF1,A1}.
Central extensions of the nonlinear sigma model with the Taub-NUT (TN)
\cite{TN} and
 Eguchi-Hanson (EH) metrics \cite{EH} and
 corresponding mass and scalar potential terms were examined in
 Refs.~\cite{IKZ,ketov}.
The domain wall
 solutions in $T^*{\bf C}P^{1}$ were also studied in the harmonic
 superspace framework, along with
 the study in ${\cal N}=1$ superfield formalism in \cite{ANNS}.

The purpose of the present paper is to investigate ${\cal N}=2$ {\it
 massive} nonlinear sigma models in the HSS approach.
We limit ourselves to the case of sigma models associated with four-dimensional HK
 multi-center metrics, because everything is drastically simplified in this case.
The component action (both its kinetic and potential parts) can be
 written in terms of the single analytic HK potential.
The resulting scalar component potential turns out to
coincide with that in \cite{GTT1}.
The general form of the
 BPS equation is derived in the multi-center case. 
As examples
 we consider sigma models associated with the TN metric and its
 generalization, the so called double Taub-NUT (DTN) 
 metric (see e.g. \cite{GORV}). The latter encompasses
both the TN and EH metrics as its two limiting cases. We
demonstrate that only in the DTN and EH cases BPS domain wall solutions exist.
The condition of the existence of SUSY vacua comes out
 as some restriction on the analytic HK potential, similarly to the ${\cal N}=1$ case
 where there arise analogous restrictions on the superpotential and K\"ahler potential.
This criterion might be useful in constructing other ${\cal N}=2$ models with
domain wall solutions.

The paper is organized as follows.
Sec.~2 contains a brief review of the general massive nonlinear
 sigma models in ${\cal N}=1$ theory.
Sec.~\ref{sec:general} is devoted to the ${\cal N}=2$ SUSY
 massive nonlinear sigma model in the HSS approach.
We start in subsection 3.1 by 
 recalling basic features of massless ${\cal N}=2$ sigma models.
The general massive sigma model with at least one isometry
 is studied in the subsection \ref{sec:general-HSS}.
In the subsection \ref{sec:multi}, we study the massive sigma model
 with $U(1)$ isometry and multi-center HK target manifolds. We show,
 in this new HSS setting, that the scalar potential is expressed
 through the norm of the Killing vector.
We also study there two particular cases with $U(1)$ isometry.
Sec.~\ref{sec:BPS} is devoted to a
 criterion which the analytic HK potential should
 satisfy to admit SUSY vacua and to a BPS equation and its solution.
The summary and concluding remarks are 
 contained in Sec. \ref{sec:concl}.
In Appendices A and B, we derive the relation between the Killing
 vectors in harmonic and ordinary spaces, as well as
 the scalar potential of the massive sigma model
 with one isometry.
In this paper, we basically follow the notation of Ref.~\cite{hss2}.
 \footnote{$diag(\eta_{\mu\nu})=(1,-1,-1,-1)$
 and $\epsilon^{12}=-\epsilon_{12}=-1$.}

\section{General massive nonlinear sigma model in $d=4,~{\cal N}=1$ theory}
The superfield action of general massive ${\cal N}=1, d=4$
 SUSY nonlinear sigma
 model is given by the most general off-shell action of
 chiral ${\cal N}=1$ superfields \cite{Zu, AF1, WessBagger}.

We denote the chiral scalar superfields, the K\"ahler
 potential for the kinetic term and
 holomorphic superpotential as
 $\Phi^i$,
 $K\left(\Phi^i, \Phi^{*i}\right)$ and ${\cal W}(\Phi)$, respectively.
The chiral superfields $\Phi^i$ comprises a complex scalar field $A^i(x)$, Weyl
 fermion $\psi^i(x)$ and an auxiliary complex scalar field $F^i(x)$:
\begin{eqnarray}
\Phi^i(y,\theta)=A^i(y)+\sqrt{2}\theta \psi^i(y)+\theta^2 F^i(y),
 ~~~~y^\mu=x^\mu+i\theta\sigma^\mu\bar{\theta}.
\end{eqnarray}
The Lagrangian is given by
\begin{eqnarray}
{\cal L}
 &=&\int d^2\theta d^2\bar{\theta} K(\Phi,\Phi^{* })
  + \left[ \int d^2\theta {\cal W}(\Phi)
  + \mbox{h.c.} \right] \nonumber \\
 &=& g_{i j^*}(A, A^*)\left(\partial_\mu A^{*j} \partial^\mu A^i +
     F^{*j}F^i
     +i \bar \psi^j \bar \sigma^\mu D_\mu \psi^i
     \right)
     + {1 \over 4}g_{ij^*,kl^*}\psi^i\psi^k\bar{\psi}^j\bar{\psi}^l
     \nonumber \\
 & & -F^i\left({1 \over 2} g_{im^*}\Gamma^{m^*}_{~j^*k^*}
       \bar{\psi}^j \bar{\psi}^k
     - {\partial {\cal W} \over \partial A^i} \right)
     -F^{*i}\left({1 \over 2} g_{mi^*}\Gamma^{m}_{~jk}
      \psi^j\psi^k
     - {\partial {\cal W}^* \over \partial A^{*i}} \right) \nonumber \\
 & & -{1 \over 2}{\partial^2 {\cal W} \over \partial A^i \partial A^j}
      \psi^i \psi^j
     -{1 \over 2}{\partial^2 {\cal W} \over \partial A^{*i} \partial A^{*j}}
      \bar{\psi}^i\bar{\psi}^j, \label{lag1}
\end{eqnarray}
where $g_{ij^*}=\partial^2 K/\partial A^i \partial A^{*j}$ is the
 K{\"a}hler metric,
 and $g_{ik^*}g^{jk^*}=\delta_i^{~j}$.
Here $D_\mu$ is the covariant derivative :
 $D_\mu \psi^i = \partial_\mu \psi^i
 + \Gamma^i_{~jk}\partial_\mu A^j \psi^k$, where
 $\Gamma^i_{~jk}=g^{il^*}\partial_{j}g_{kl^*}$ is the holomorphic part of the Levi-Civita
 connection. We denote the derivation by
 $g_{jl^*,k^*} \equiv {\partial g_{jl^*} \over \partial A^{*k}}$.
Equations of motion for $F^i$ are
\begin{eqnarray}
 F^i=\frac{1}{2}\Gamma^i_{~jk}\psi^j\psi^k
     - g^{ij^*}{\partial {\cal W}^* \over \partial A^{*j}}.
 \label{aux1}
\end{eqnarray}
Substituting (\ref{aux1}) into (\ref{lag1}), we obtain
\begin{eqnarray}
{\cal L}
&=&
 g_{i j^*}(A, A^*)\left(\partial_\mu A^{*j} \partial^\mu A^i
 +i\bar \psi^j \bar \sigma^\mu D_\mu \psi^i
 \right)+{1 \over 4}R_{ij^*kl^*}\psi^i\psi^k\bar{\psi}^{j}\bar{\psi}^l
 \nonumber \\
&&
 - \frac{1}{2}D_iD_j {\cal W} \psi^i \psi^j
 - \frac{1}{2}D_{i^*}D_{j^*}{\cal W}^* \bar{\psi}^i \bar{\psi}^j
 - V (A, A^*)
\end{eqnarray}
where $R_{ij^*kl^*}$ is the curvature tensor defined by
 $R_{ij^*kl^*} =
 g_{ij^*,kl^*}-g_{mn^*}\Gamma^m_{~ik}\Gamma^{n^*}_{~j^*l^*}$
 and $D_i$ is the covariant derivative in the target space :
\begin{eqnarray}
D_i {\cal W} &=& {\partial \over \partial A^i} {\cal W}, \\
D_i D_j {\cal W} &=& {\partial^2 \over \partial A^i \partial A^j}{\cal W}
              -\Gamma^k_{~ij}{\partial \over \partial A^k}{\cal W}.
\end{eqnarray}
The scalar potential $V (A, A^*) $ is given by
\begin{equation}
V (A, A^*) = g^{i j^*} {\partial {\cal W} \over \partial A^{i}}
 {\partial {\cal W}^* \over \partial A^{*j}}
 =
 g_{i j^*} F^{i}  F^{*j}.
 \label{eq:scalarpotential}
\end{equation}

From (\ref{eq:scalarpotential}), it is easy to find out
 the SUSY vacuum condition.
The condition of the SUSY vacuum is the vanishing of the scalar potential
\begin{equation}
0=V (A, A^*) =
g^{i j^*} {\partial {\cal W} \over \partial A^{i}}
{\partial {\cal W}^* \over \partial A^{*j}}
=
g_{i j^*} F^{i}  F^{*j} .
\label{eq:SUSYcondition}
\end{equation}

To simplify things, let us consider the case of the nonlinear
 sigma model
 with only one chiral scalar superfield $\Phi$.
We find that there are two cases when the SUSY vacuum exists
 in such a nonlinear sigma model, namely, if
\begin{enumerate}
 \item there is a stationary point of the superpotential, such that the
 K\"ahler metric is not vanishing at this point
\begin{equation}
{\partial {\cal W} \over \partial A} = 0,
\quad  {\rm and} \quad
g_{A A^*} \not= 0
\quad \left( \; g^{A A^*} \not= \infty \; \right),
\label{eq:stationary_superpotential}
\end{equation}
\item there is a singularity of the K\"ahler metric, such that the derivative of the
superpotential is not singular at this point
\begin{equation}
g_{A A^*} = \infty
\quad \left( \; g^{A A^*} = 0 \; \right),
\quad {\rm and} \quad
{\partial {\cal W} \over \partial A} \not= \infty .
\label{eq:singularity_kahlermetric}
\end{equation}
\end{enumerate}

Note that the vanishing of $F$ term ($F^i=0$) is
neither necessary nor sufficient for SUSY to be
unbroken in the nonlinear sigma model,
since the K\"ahler metric in eq. (\ref{eq:SUSYcondition})
can possess zeros and/or singularities.

Next, let us consider the BPS equation.
In the following we examine only the bosonic part since
 we are now interested in the non-trivial bosonic configuration.
The BPS equation can be derived from the energy minimum condition.
Assuming that there exists a non-trivial configuration along
 the spatial $y$ axis,
 the energy of the general nonlinear sigma model (per unit area)
 is expressed as
\begin{eqnarray}
E &=& \int dx_2 \left(g_{ij^*}\partial_2 A^{*j} \partial_2 A^i
      +g^{ij^*}{\partial {\cal W} \over \partial A^i}
               {\partial {\cal W}^* \over \partial A^{*j}} \right)
      \nonumber \\
  &=& \int dx_2 \left\{ g_{ij^*}\left(\partial_2 A^i - e^{i\delta}
                        g^{ik^*} {\partial {\cal W}^* \over \partial A^{*k}}
                                \right)
                                \left(\partial_2 A^{*j} - e^{-i\delta}
                        g^{lj^*} {\partial {\cal W} \over \partial A^l}
                                \right)
             + 2 \partial_2 {\rm Re} (e^{-i\delta} {\cal W}) \right\}
      \nonumber \\
  &\ge& 2 \int dx_2 \partial_2 {\rm Re} (e^{- i \delta} {\cal W}),
\end{eqnarray}
where $\delta$ is a phase factor.
If we impose the most restrictive bound, a phase of the superpotential
 can be taken as
\begin{eqnarray}
 \delta = arg({\cal W}(y=\infty)-{\cal W}(y=-\infty)).
\end{eqnarray}
When the bound is satisfied, the following equation must hold
\begin{eqnarray}
 \partial_2 A^i = e^{i\delta} g^{ij^*}{\partial {\cal W}^*
  \over \partial A^{*j}}
 \label{eq:BPSeq}
\end{eqnarray}
where both sides are evaluated at classical fields.
This equation is called the BPS equation, and we thus find that
 the BPS equation involves both the K{\"a}hler
 potential and the superpotential. So the existence of non-trivial solutions
 to the BPS equation is encoded in the structure of these entities.

This formulation is useful when we are interested in massive
 nonlinear sigma models having
 domain wall solutions.
To possess a domain wall solution, the theory should have at least
 two discrete vacua.
The conditions (\ref{eq:stationary_superpotential}) and
 (\ref{eq:singularity_kahlermetric}) can then be used to set up such a model.
For instance, in Ref. \cite{NNS1},
 the model corresponding to the choice $g_{AA^*}\equiv K_{AA^*}=1/|1-A^2|$ was given as a simple example.
Once the model is specified, one can easily obtain the domain wall
 solution using (\ref{eq:BPSeq}).

%
%
\section{General massive nonlinear sigma model in the harmonic superspace}
\label{sec:general}
First, we study the general massive nonlinear sigma model with at
 least one triholomorphic (i.e., commuting with
 supersymmetry) $U(1)$ isometry.
The presence of such an isometry is necessary if one
 wishes to gain the mass (and/or scalar potential) terms.
We shall not specify how this isometry is realized.
Next we examine the particular case of the four-dimensional
 target HK space.
In this case, requiring the theory to have an $U(1)$ isometry
 implies that the corresponding HK metric
 falls into the multi-center class \cite{GORV}.
As was shown in \cite{GORV}, using some coordinate transformation,
 this $U(1)$ isometry can always be cast in the form in which
 it is realized as some phase or purely shift transformation of the
 coordinates of the HK manifold.
For this case we shall
 demonstrate that the scalar potential is given
 by the square of the isometry Killing vector,
 in accord with the result of \cite{GTT1}.

\subsection{HK sigma model in HSS: the general massless case}

First, we consider the action of the general massless nonlinear
sigma model in the HSS approach.

The HSS action for a general nonlinear ${\cal N}=2, d=4$ supersymmetric
sigma model which yields in the bosonic sector a sigma model with $4n$ dimensional HK
target space is just the general superfield action of $n$ hypermultiplets.
In the HSS formalism, the hypermultiplet is described by an analytic superfield
 $q_a^{+}$ ($a=1,\ldots ,n$ is a flavor index of fundamental representation of $Sp(n)$)
 which is a function given on the harmonic analytic ${\cal N}=2$ superspace
\begin{eqnarray}
 \{\zeta_A,u_i^{\pm}\} \equiv \{
 x^\mu_A=
 x^\mu - 2 i\theta^{(i}\sigma^\mu\bar{\theta}^{j)}u^+_{(i}u^-_{j)},~
 \theta^+=\theta^iu_i^+,~\bar{\theta}^+=\bar{\theta}^i u_i^{+},~
 u_i^{\pm} \}, \label{ahss}
\end{eqnarray}
where
 the coordinates $u^{+i},u^{-i},u^{+i}u_{i}^-=1,~i=1,2$
 \footnote{In what follows, $a,\ldots,f$ stand for the $Sp(n)$ indices 
 and $i,j,\ldots$ for the
 $SU(2)_R$ indices, respectively.} are the $SU(2)_R/U(1)$ harmonic variables \cite{hss1,hss2}.

Exploiting the target space reparameterization covariance,
 the general action can
be cast in the form \cite{hss2}
\begin{eqnarray}
S=\frac{1}{2}\int d\zeta_A^{(-4)}du\left[q_a^+D^{++}q^{a+}
 +L^{+4}(q_a^+,u_i^\pm) \right], \label{hsfaction}
\end{eqnarray}
where $d\zeta_A^{(-4)} du = d^4x_A d^2\theta^+d^2\bar{\theta}^+du$ is the
 measure of integration over analytic superspace (\ref{ahss}),
 $D^{++}$ is the harmonic covariant derivative
 defined as
\begin{eqnarray}
D^{++}=\partial^{++}-2i\theta^+\sigma^\mu{\bar{\theta}}^+\partial_\mu\,,
 \,\,\,\, \partial^{++}=u^{+i}{\partial \over \partial u^{-i}}
 \label{covariant}
\end{eqnarray}
and $L^{+4}(q_a^+,u_i^{\pm})$ is the analytic HK potential.
The analytic superfield $q_a^{+}$
 can be expanded as
\begin{eqnarray}
 q_a^+(\zeta_A,u_i^\pm) &=& F_a^+ + \sqrt{2} \theta^+ \psi_a
                      + \sqrt{2} {\bar{\theta}}^+ {\bar{\varphi}}_a
                      + i\theta^+ \sigma^\mu {\bar{\theta}}^+ A_{a\mu}^-
                      + \theta^+ \theta^+ M_a^-
                      + {\bar{\theta}}^+{\bar{\theta}}^+ N_a^-
                        \nonumber \\
                  & & + \sqrt{2} \theta^+\theta^+{\bar{\theta}}^+
                        {\bar{\chi}}_a^{--}
                      + \sqrt{2} {\bar{\theta}}^+ {\bar{\theta}}^+
                        \theta^+ \xi_a^{--}
                      + \theta^+ \theta^+ {\bar{\theta}}^+ {\bar{\theta}}^+
                        P_a^{(-3)},
                      \label{eq:exp1}
\end{eqnarray}
 and it satisfies the reality condition
\begin{eqnarray}
 \widetilde{q_a^+}=\Omega^{ab}q_b^+\,.
\end{eqnarray}
Here
 $\Omega^{ab}$ is the skew-symmetric constant $Sp(n)$ metric,
 $\Omega^{ab}\Omega_{bc} = \delta^a_c$, and
 ``$\sim$'' denotes a (pseudo) conjugation which is
 the product of the complex conjugation (denoted by ``$ - $'')
 and the star (pseudo) conjugation \cite{hss2}.
The action of the ``$\sim$'' conjugation on $\zeta_A$ and $u_i^\pm$ is defined as
\begin{eqnarray}
\widetilde{x_A^\mu}=x_A^\mu,~\widetilde{\theta^+}=\bar{\theta}^+,
~\widetilde{\bar{\theta}^+}=-\theta^+,~\widetilde{u_i^{\pm}}=u^{\pm i},
~\widetilde{u^{\pm i}}=-u^{\pm}_i.
\end{eqnarray}
In what follows, we shall frequently omit the $Sp(n)$,~$SU(2)_R$ and space-time indices of the arguments in the
 analytic functions, e.g. write $f=f(q_a^+,u_i^\pm)=f(q,u)$.

The action \p{hsfaction} is assumed to be invariant under the following isometry transformation
\footnote{In general, \p{hsfaction} is not obliged to respect any extra symmetry except for ${\cal N}=2$ SUSY.}
\begin{eqnarray}
 \delta q^{a+} = \varepsilon\, \lambda^{a+}(q, u), \label{1a}
\end{eqnarray}
provided that $\lambda^{a+}(q, u)$ satisfies the equations
 ($\partial_{a+}=\partial/\partial q^{a+}$)
\begin{eqnarray}
&& \lambda^{a+} = {1\over 2}\Omega^{ab}
                  \partial_{b+}\Lambda^{++}, \label{1b} \\
&& \partial^{++}\Lambda^{++} -{1\over 2}\Omega^{ab}
                  \partial_{a+}L^{+4}\partial_{b+}\Lambda^{++} = 0\,.
                  \label{1c}
\end{eqnarray}
In eq. \p{1a}, $\varepsilon$ is a group parameter.
The quantities $\lambda^{a+}$ and $\Lambda^{++}$
 are referred to as the superfield Killing vector and
 Killing potential, respectively.
In what follows we shall need eqs. \p{1a} - \p{1c} only in the limit when
 all fermions are discarded, which amounts to the reduction
 $q^+ \rightarrow F^+$.

From now on, we neglect all fermionic fields and deal with the bosonic
component action. Both fermionic and bosonic components in \p{eq:exp1} contain
infinite sets of auxiliary fields coming from the harmonic expansions.
In order to obtain the action in terms of $4n$ physical bosonic fields only,
we should eliminate the relevant auxiliary fields by solving their 
 algebraic (i.e., kinematical)
equations of motion. Therefore, as the basic steps towards
 the final sigma model action we
should single out the kinematical part of the equations of motion following from
\p{hsfaction} (with all fermions being discarded) and solve these equations.

Substituting the bosonic part of Grassmann expansion (\ref{eq:exp1})
 into the action \p{hsfaction}, and integrating over Grassmann coordinates, 
 we obtain the bosonic action in the form
\begin{eqnarray}
 S_{bos}&=&\displaystyle \int d^4 x_A du {\Bigg \{}
  \frac{1}{4}A_\mu^{a-}
  \left[\left(D^{++}\delta_{a}^{~b}
   -{1 \over 2}\partial_{a+}\partial_{c+}\Omega^{cb}L^{+4}\right)
   A_b^{-\mu}-4\partial^\mu F_a^+ \right]
  \nonumber \\
 &&-M^{a-}\left(D^{++}\delta_a^{~b}
   -{1 \over 2}\partial_{a+}\partial_{c+}\Omega^{cb}L^{+4}\right)N_{b}^-
   -P^{a(-3)}\left(D^{++}F_a^+ -\frac{1}{2}\partial_{a+}L^{+4}
 \right) {\Bigg \}}. \label{compac1}
\end{eqnarray}
Here $L^{+4}=L^{+4}(F,u)\equiv L^{+4}(q,u)$ at $\theta=0$.
Equation of motion for $F^+_a$ is
\begin{eqnarray}
&&D^{++}F^{+}_a(x,u) - {1\over 2}\partial_{a+}L^{+4}(F, u)=0\,.\label{2a}
\end{eqnarray}
Here $D^{++}$ coincides with a partial harmonic derivative
$\partial^{++}$ which acts on the harmonic arguments of the component fields as well as on the harmonics
appearing explicitly.
We denote the harmonic derivative in \p{2a} by $D^{++}$
 in order to distinguish it, e.g., from the partial derivative in \p{1c} which acts only on the
 explicit harmonics in $\Lambda^{++}(F,u)$ and not on the harmonic arguments of $F = F(x,u)$.
 We reserve the notation $\partial^{++}$ just for this latter derivative.

The derivative $D^{++}$, when applied to an arbitrary scalar function $G(F,u)$, yields:
\beq
 D^{++}G = \partial^{++}G + D^{++}F^{a+}\partial_{a+}G = \partial^{++}G +
 {1\over 2}\Omega^{ab}\partial_{b+}L^{+4}\partial_{a+}G\,, \label{DG}
\eeq
where we used eq. \p{2a}.
Defining
\begin{eqnarray}
{\cal D}^{++}G_a=D^{++}G_a-{1 \over 2}\partial_{a+}\partial_{b+}L^{+4}G^b
 \label{5a},
\end{eqnarray}
equations of motion for $A_{a\mu}^-$,
 $M_a^-$ and $N_a^-$ can be rewritten as
\begin{eqnarray}
&& {\cal D}^{++}A_{a\mu}^{-}-2\partial_\mu F_a^+ = 0\,,\label{eq4} \\
&& {\cal D}^{++}M_{a}^-
 = {\cal D}^{++}N_{a}^- = 0\,\label{eq5}.
\end{eqnarray}
The remaining equation, which comes from the variation with respect to
 $F_a^+$, is dynamical, and we will not use it in the following
 (it can be reproduced in the end by varying the eventual action with respect to the dynamical
 fields $f^{ai}(x)$ to be defined below).

After substituting eqs. \p{2a},~\p{eq4} and \p{eq5} back into \p{compac1},
 the action is drastically simplified
\begin{eqnarray}
S_{bos} =\displaystyle\int d^4x_A du\left(
 -\displaystyle\frac{1}{2}A_\mu^{a-}\partial^\mu F_a^+
 \right).
 \label{compac2}
\end{eqnarray}
Note that the harmonic fields $F_a^+$ and $A_{a\mu}^-$ are
 still subject to
 the constraints \p{2a} and \p{eq4} and they include infinite sets of
 auxiliary fields.
Solving eq. \p{2a}, one can express $F^{+}_a(x_A,u)$ as
 $F^{+}_a = F^{+}_a(f^{ai}, u)$ where
 $f^{ai}(x)$ are the standard HK target space coordinates.
We shall refer to
\beq
f^{a\pm} = f^{ai}(x)u^{\pm}_i
\eeq
as the ``central basis'' HK coordinates and to $F^{a+}$ and $F^{a-}$ related by
\beq
D^{++}F^{a-} = F^{a+} \label{F-F+}
\eeq
as the ``analytic basis'' HK coordinates \cite{hss3,hss2}.
A more detailed explanation of this nomenclature can be found
 in \cite{hss3,hss2}.
Given the solution $F^{+}_a=F_a^+(f^{ai},u)$, the field $A_{a\mu}^-$ can be
expressed from \p{eq4} as $A_{a\mu}^-=A_{a\mu}^-(f^{ai},u)\,$.
Substituting these solutions into the action results in the final sigma model
action for $f^{ai}(x)$.

In solving eqs. \p{2a}, \p{eq4} and \p{eq5},
 it is convenient to make use of the
 one-to-one correspondence between the HK sigma models and the
 geometric construction
 of HK manifold in the harmonic space \cite{hss3,hss2}.
In the latter formulation, the standard constraints of the HK geometry are
interpreted as the conditions of harmonic analyticity.
This allows one to solve defining constraints of the HK geometry in terms of
two unconstrained analytic potentials one of which proves to be pure gauge.
The remaining potential encodes all the information about the given HK
 manifold,
 in the sense that all the relevant geometric objects, i.e.,
 connections, vielbeins and metric,
 can be expressed in terms of this potential.
We call this geometric approach the non-Lagrangian
 one, in contrast to the Lagrangian approach to which we adhere in
 this paper and in which
 the metric and other geometric quantities of HK geometry appear in
 the ${\cal N}=2, d=4$
 supersymmetric sigma model type action as the result of solving
 equations of motion for
 an infinite tower of auxiliary fields contained in $q^{a+}$.
As was shown in \cite{hss3,hss2}, both approaches
 are in fact equivalent to each other.
In particular, the unconstrained potential in the
 non-Lagrangian approach corresponds to the analytic HK potential
 in ${\cal N}=2$ nonlinear sigma model.
Using the one-to-one correspondence between these two approaches,
 as well as the differential
 geometry techniques of Refs. \cite{hss3,hss2},
 it is easy to check that the solution of eqs. \p{eq4}
 and \p{eq5} is
\begin{eqnarray}
&& A_{\mu}^{a-} = 2E_{bi}^{a-}\partial_\mu f^{bi}, \label{sol1}\\
&& M_a^-=N_a^-=0, \label{sol2}
\end{eqnarray}
where $E_{bi}^{a-}$ is one of the two central basis vielbeins
 $E_{bi}^{a\pm}$ from which the HK metric is constructed.
They satisfy the relations \cite{hss2}
\beq
{\cal D}^{++}E^{a+}_{bi} = 0\,, \quad
 E^{a+}_{bi} = - {\cal D}^{++} E^{a-}_{bi} =
 -\partial_{bi}F^{a+}\,,\label{8a}
\eeq
which can be deduced based upon eq. \p{2a} (for instance,
 the first one is
 proved by applying $\partial/\partial f^{bi}$ to \p{2a}).
Substituting \p{sol1} and \p{sol2} into \p{compac2}, we obtain
\begin{eqnarray}
 S_{bos}={1 \over 2} \int d^4 x_A \,\, g_{ai,bj}(x_A) \,
 \partial^\mu f^{ai}\partial_\mu f^{bj}, \label{last}
\end{eqnarray}
where $g_{ai,bj}$ is the HK target space metric defined by
\begin{eqnarray}
g_{ai,bj}=\Omega_{cd}(E_{ai}^{c-}E_{bj}^{d+}-E_{ai}^{c+}E_{bj}^{d-}).
 \label{metric-def}
\end{eqnarray}
It is easy to show that this metric is $u$ independent,
 $g_{ai,bj}=g_{ai,bj}(x_A)$. This can be checked utilizing
 eq. \p{8a}, keeping in mind that the $Sp(n)$ connections drop
 out altogether due to the contraction of the $Sp(n)$ indices.

We briefly sketch
 the basic steps
 leading from the action \p{compac2} to the final action \p{last}.
Firstly, we use in \p{compac2} the relation
$$
\partial^\mu F^{a+} = \partial^\mu f^{ck}\partial_{ck}F^{a+} =
-\partial^\mu f^{ck}E^{a+}_{ck}\,,
$$
after which the integrand in \p{compac2} is reduced to
$$
\Omega_{ad} E_{bi}^{a-}E^{d+}_{ck}\partial_\mu f^{bi}
\partial^\mu f^{ck}\,.
$$
Then we take into account that the second piece
 in \p{metric-def} is reduced to the
 first one under the $u$-integral: one should use eq. \p{8a},
 take into account that the
 $Sp(n)$ connections drop out due to the contractions
 of the $Sp(n)$ indices, and finally integrate by parts with respect
 to $D^{++}$.
After these manipulations the last
 expression is reduced, modulo a total harmonic derivative, just
 to the metric term in \p{last}.

To close the discussion of the massless nonlinear sigma model,
we collect some useful formulas related to the $U(1)$ isometry \p{1a}
projected on the bosonic sector. The basic HSS Killing potential equation \p{1c}
on the surface of the kinematic equation \p{2a} amounts to the `conservation law':
\beq
D^{++}\Lambda^{++}(F,u) = 0\,. \label{4a}
\eeq
In the central basis, eq. \p{4a} implies
\beq
\Lambda^{++}(f, u) = \Lambda^{ik}(f)u^+_iu^+_k\,.\label{9a}
\eeq
Using \p{4a} and \p{5a}, it is also easy to find
\beq
{\cal D}^{++} \lambda^{a+}(F,u) = 0\,. \label{36a}
\eeq
{}From eqs. (\ref{1a}) - (\ref{1c}) and also using \p{9a},
we find the following central basis form of \p{1a}
\beq
\delta F^{a+} = \varepsilon\, \lambda^{a+}(F,u) = \varepsilon\, {1\over 2}\Omega^{ab}
 \partial_{b+}f^{ck}\partial_{ck}\Lambda^{ij}(f)u^+_iu^+_j =
\varepsilon\, \partial_{ck}F^{a+}k^{ck}(f)
 = - \varepsilon\, E^{a+}_{ck}k^{ck}\,. \label{6a}
\eeq
Here $k^{ck}(f)$ is the ordinary Killing vector,
\beq
\delta f^{ck} = \varepsilon\, k^{ck}(f)\,.\label{7a}
\eeq
The identity \p{6a} allows one to find the explicit expression for
 $k^{ai}(f)$ in terms of the central basis Killing potential
 $\Lambda^{(ij)}(f)$ (see Appendix A).

%
%
\subsection{General massive HK sigma model in HSS}
\label{sec:general-HSS}

Next we consider the general massive deformation of the HSS
 $q^+$ Lagrangian.
Suppose we are given a $q^{+}$ action possessing an isometry.
Then we assign to $q^{+}$ a dependence on the central charge coordinate
 $x^5$, such that
 $\partial/\partial x^5$ can be identified with the Killing vector
 of the isometry
\begin{eqnarray}
\frac{\partial}{\partial x^5}\, q^{a+} = m\, \lambda^{a+}(q, u)\;\;
 \label{11a}
\end{eqnarray}
where $m$ is a mass parameter which, for simplicity, is taken to be real.
Correspondingly, the harmonic covariant derivative \p{covariant}
 acquires the central charge term \cite{Ohta}:
\beq
D^{++} \quad \rightarrow \quad D^{++} +
i\left[(\theta^+)^2- (\bar\theta^+)^2\right]\frac{\partial}{\partial x^5}\,,
\eeq
and the action \p{hsfaction} is modified as
\beq
S=\frac{1}{2}\int d\zeta^{(-4)}_A du \left(
 q^+_a D^{++}q^{a+} + L^{+4}(q,u)
 + im \left[(\theta^+)^2- (\bar\theta^+)^2\right]q^+_a\lambda^{a+}
 \right). \label{10a}
\eeq
Similarly to the massless case,
 substituting the Grassmann expansion of the harmonic superfield
 (\ref{eq:exp1}) with suppressed fermions into \p{10a},
 we obtain the bosonic action in the following form
\begin{eqnarray}
 S_{bos}&=&\int d^4 x_A du {\Bigg (}
 \frac{1}{4}A_\mu^{a-}
 \left({\cal D}_{~~a}^{++b}A_b^{-\mu}-4\partial^\mu F_a^+ \right)
 -M^{a-}{\cal D}_{~~a}^{++b}N_{b}^- \nonumber \\
&-&\,P^{a(-3)}\left(D^{++}F_a^+
 -{1 \over 2} \partial_{a+} L^{+4}
  \right)+im(N^{a-}\partial_{a+}(F_b^+\lambda^{b+})
   -M^{a-}\partial_{a+}(F_b^+\lambda^{b+}))
 {\Bigg )}. \label{B100}
\end{eqnarray}
The corresponding equations of motion read
\begin{eqnarray}
&&{\cal D}^{++}M_{a}^- - im\,\partial_{a+}(F^+_b\lambda^{b+}) = 0\,,
 \label{13a}\\
&&{\cal D}^{++}N_{a}^- + im\,\partial_{a+}(F^+_b\lambda^{b+}) = 0\,,
 \label{13b}
\end{eqnarray}
along with \p{2a} and \p{eq4}.
It is worth emphasizing that the equations of motion for 
 $M_a^-$ and $N_a^-$ are modified as compared to \p{eq5}
 due to the mass deformation, while the equations \p{2a} 
 and \p{eq4} for $F_a^+$ and $A_{a\mu}^-$
 are not modified. 
As in the massless case, these equations 
 serve to express
 infinite sets of auxiliary fields collected in the involved quantities
 in terms of the central basis HK coordinate $f^{ai}(x)$ and 
 its $x$-derivative.

After substituting these kinematical equations of motion into (\ref{B100}),
 the bosonic component action acquires the simple form
\begin{eqnarray}
S_{bos} =\displaystyle\int d^4x_A du \left[
 -\displaystyle\frac{1}{2}A_\mu^{a-}\partial^\mu F_a^+
 -\frac{im}{2}(M^{a-}-N^{a-})\partial_{a+}(F_b^+\lambda^{b+})\right].
 \label{component2}
\end{eqnarray}
The harmonic fields in (\ref{component2}) are still solutions of
 \p{2a}, \p{eq4}, \p{13a} and \p{13b}.
Since we have already solved the equations \p{2a} and \p{eq4}
 while studying the
 massless case,
 the remaining equations to be solved are \p{13a} and \p{13b}.
Here we present the general form of the solution for $M^{a-}, N^{a-}$ (some details of
the derivation are given in Appendix B):
\begin{eqnarray}
M^{a-} = -N^{a-}
 =im \left({\cal F}^-_b\partial^a_+\lambda^{b+} + k^{ci}\left[\partial_{ci} {\cal F}^{a-}
    - 2\,E^{a-}_{ci}\right] \right)\,.\label{17a}
\end{eqnarray}
Here ${\cal F}_a^{-}$ is defined by
\beq
F^+_a = {\cal D}^{++}{\cal F}^-_a = D^{++}{\cal F}^-_a
 - {1\over 2}\partial_{a+}\partial_{b+}L^{+4}{\cal F}^{b-}\,.
\label{15a}
\eeq

Substituting (\ref{sol1}) and (\ref{17a}) into (\ref{component2}),
 we obtain,
\begin{eqnarray}
S_{bos}&=& S_{kin} + S_{pot} = {1 \over 2} \int d^4 x_A \,\, g_{ai,bj}(x_A) \,
 \partial^\mu f^{ai}\partial_\mu f^{bj}
 -\int d^4 x_A V(f^{ai}), \label{finact} \\
V(f^{ai})&=&
     m^2 \int du \, \partial_{a+}(F_b^+\lambda^{b+})\Omega^{ad}
     \left[ {\cal F}_e^{-}\partial_{d+}\lambda^{e+}+k^{ci}(\partial_{ci}{\cal F}_d^-
     -2\Omega_{de}E_{ci}^{e-})\right]. \label{finalscalar}
\end{eqnarray}
The kinetic sigma model term in \p{finact} has the same form as the massless bosonic
 action \p{last}.
Note that the potential in the generic case still displays a
  harmonic dependence while the
 kinetic term does not depend on the harmonic variables.
The genuine scalar potential in $x$-space is
 obtained after performing the integration over harmonics.
In Appendix B, we give the sufficient condition under which this
integration can be explicitly performed.

%
%
\subsection{Massive HK sigma model in HSS: a multi-center case}
\label{sec:multi}
In the previous subsection we derived the component action
 of the general massive HK nonlinear sigma model with at least
 one triholomorphic isometry.
We did not specify the precise realization of this isometry.
We obtained the kinetic term of the nonlinear sigma model
 which has exactly the form prescribed in Refs. \cite{Zu,AF1}.
However, in the general case the harmonic integral in the scalar potential
 cannot be computed in a simple way. Fortunately, in the case
 of four-dimensional HK manifolds the situation is simplified
 radically
 due to the theorem \cite{Gibbons,GORV} claiming that any 4-dimensional HK metric
 with at least one $U(1)$
 triholomorphic isometry falls into the class of Gibbons-Hawking multi-center
 metrics \cite{GiHa}.
Moreover, it can be shown (see \cite{hss2} and refs. therein)
 that the HK potentials for such metrics can always be brought to
 the form $L^{+4}_{mc} = L^{+4}(u^+\cdot q^+, u)$ where $u^+\cdot q^+ = u^{+a}q^+_a$ and the
 isometry is realized as the shift $q^{+a} \rightarrow q^{+a} + \varepsilon\, u^{+a}$.
As a result, the computation of the potential is drastically simplified.

In the present case we have
\beq
\lambda^{a+} = u^{+a}\,,  \quad \frac{\partial}{\partial x^5}\, q^{a+}
 = m\,u^{+a}\,. \label{19a}
\eeq
In this particular case, the Lagrangian in \p{component2} can be rewritten
 as follows
\begin{eqnarray}
{\cal L}&=&{\cal L}_{kin}+{\cal L}_{pot}, \label{Lmc}\\
{\cal L}_{kin}&=&-{1 \over 2} \int du A_\mu^{a-}\partial^\mu F_a^+,
                \label{Lmckin}\\
{\cal L}_{pot}&=&-im \int du M^{a-}u^+_a,
                \label{61a}
\end{eqnarray}
where we used $M_a^-=-N_a^-\,$, which follows from \p{13a} and
 \p{13b}.

Our purpose is to derive the component action of physical bosons in $x$-space from the
$(x,u)$-space action \p{Lmc}.
One of the ways to obtain it is to substitute \p{19a} into
 the general formula \p{finact}.
However, it is easier to proceed directly
 by solving eqs. \p{2a}, \p{eq4}, \p{13a} and \p{13b}.
We carry out this in two steps.
First, we solve the equations of motion \p{2a} and \p{eq4},
 and derive the kinetic term.
As a result of solving these equations, $F_a^{+}$ and $A_{a}^{\mu-}$ 
 are expressed in terms
of the dynamical physical fields $f^{ai}$.
It turns out that it is actually enough to solve 
 equation for $A_{a}^{\mu -}$ partially, as distinct
from  the equation for $F_a^+$ which should be solved exactly.
Secondly, we solve the equations
 \p{13a} and \p{13b}.
These solutions are needed to derive the scalar potential.
We will see that the scalar potential is expressed in terms of
 the analytic HK potential after substituting the solutions into
 \p{61a}.

The equations of motion \p{2a} and \p{eq4} are written in the considered particular case as
\begin{eqnarray}
&& D^{++}F_a^{+}-u_a^{+}L^{+2} = 0\,, \label{eq1t} \\
&& D^{++}A_a^{\mu-}-u_a^+(u^+\cdot A^{\mu-})-2\partial^\mu F_a^+ = 0\,,\label{eq4t}
\end{eqnarray}
where
\begin{eqnarray}
&L^{+2}(u^+\cdot F^+,u) =-\displaystyle
               \frac{1}{2}\frac{\partial L^{+4}}{\partial (u^+\cdot F^+)},&
               \nonumber \\
&L(u^+\cdot F^+,u) =-\displaystyle
               \frac{1}{2}\frac{\partial^2 L^{+4}}{\partial (u^+\cdot F^+)^2}\,.&
               \label{22a}
\end{eqnarray}
First we solve \p{eq1t}.
Substituting the following ansatz
\begin{eqnarray}
F^{a+}=f^{ai}u^+_i+v^{a+}(f^{ai},u)
\end{eqnarray}
into \p{eq1t}, we obtain
\begin{eqnarray}
\partial^{++}v^{a+}(f^{ai},u)=u^{+a}L^{+2}. \label{eq1t3}
\end{eqnarray}
Up to a gauge freedom, $v^{a+}(f)$ can be written as
\begin{eqnarray}
v^{a+}(f^{ai},u)=u^{+a}v(f^{ai},u).
\end{eqnarray}
Indeed,
\begin{eqnarray}
u^{+} F^{+} = u^{+a}(f_a^{~i}u_i^+ + u_a^+ v(f^{ai},u))
 = u^{+a}f_a^{~i} u_i^+.
\end{eqnarray}
Thus, eq. \p{eq1t3} amounts to
\begin{eqnarray}
\partial^{++} v(f^{ai},u)=L^{+2}(u^+\cdot f^+,u). \label{eq1t4}
\end{eqnarray}
Using the harmonic Green function \cite{hss2},
 we obtain the general solution of this equation as
\begin{eqnarray}
v(f^{ai},u)=\int dw\, \frac{u^+\cdot w^-}{u^+\cdot w^+}\,L^{+2}(w^+\cdot  f^+,w)\,.
\end{eqnarray}
To be convinced that this is indeed a solution,
 one should substitute it into \p{eq1t4} and
 use
\begin{eqnarray}
\partial_u^{++}\left(\frac{u^+\cdot w^-}{u^+\cdot w^+}\right)=\delta^{(2,-2)}(u,w).
\end{eqnarray}
Thus, we find the final form of the solution of \p{eq1t} to be
\begin{eqnarray}
F^{a+}=f^{ai}u_i^+
 + u^{+a}\int dw\, \frac{u^+\cdot w^-}{u^+\cdot w^+}\,L^{+2}(w^+\cdot f^+,w) \label{sol-eq1}
\end{eqnarray}
which will be used for computing the kinetic term.

Next, we partially solve eq. \p{eq4t}.
Multiplying \p{eq4t} by $u_a^{+}$ and $u_a^{-}$, we obtain
\begin{eqnarray}
&&D^{++}(u^+\cdot {\tilde{A}}_\mu^{-})=0\,, \label{proj1} \\
&&D^{++}(u^-\cdot \tilde{A}_\mu^{-})-
 (u^+\cdot {\tilde{A}}_\mu^-)(1-L)+2u^{+a}\partial_\mu f_a^{~i} u_i^-L
 +2 \partial_\mu v(f^{ai},u) = 0\,, \label{proj2}
\end{eqnarray}
where
\begin{eqnarray}
\tilde{A}_a^{\mu-}=A_a^{\mu-}-2\partial^\mu f^{~i}_a u_i^-.
 \label{defA}
\end{eqnarray}
Eq. \p{proj1} implies that $B^\mu(x) \equiv (u^+\cdot
 {\tilde{A}}^{\mu-})$ does not depend on the harmonics.
Substituting this into \p{proj2} and taking the harmonic 
 integral of the l.h.s. of \p{proj2}, we find
\begin{eqnarray}
 B^\mu(x)=-\frac{2}{1+V_0}
           \partial^\mu f^{ai} V_{ai}\,,\label{psol}
\end{eqnarray}
where
\begin{eqnarray}
V_{ai}&=&\displaystyle\int du\,u_a^+u_i^- L(u^+\cdot f^+,u), \\
V_0&=&\epsilon^{ai}V_{ai}=-
 \int du\,L(u^+\cdot f^+,u).
 \label{relation1}
\end{eqnarray}
Here, we have used the property
\beq
\int du\, v(f, u) =
 \int du dw\, \frac{u^+\cdot w^-}{u^+\cdot w^+}\,L^{+2}(w^+\cdot f^+,w) = 0\,,\label{vanv}
\eeq
which can be proved by representing $u^{+}_i$ in the numerator of the integrand
as $u^{+}_i = \partial^{++}_u u^{-}_i$, integrating by parts with
 respect to $\partial^{++}_u$ and using the properties
$$
(u^-\cdot w^-)\,\delta^{(1,-1)}(u, w) = 0\,
$$
and
\begin{eqnarray*}
 \int du\, \partial^{++}f^{(q)}(u)=0 \,
\end{eqnarray*}
where $q$ is a $U(1)$ charge.
From \p{defA} and \p{psol}, we obtain
\begin{eqnarray}
A_{\mu a}^{-}= -\left[ u^+_a (u^-\cdot A^-_\mu)
 + 2\,u^-_a\left( \partial_\mu f^{bj}u^+_b u_j^-
 + \frac{1}{1+V_0}\partial_\mu f^{bj}\,V_{bj}\right)\right]\,.
 \label{sol-eq4}
\end{eqnarray}

Now we are ready to compute the kinetic term.
As already mentioned, in order to compute the metric, there is no
 need to explicitly solve eq. \p{proj2}
 for the remaining unknown
 $(u^-\cdot A^-_\mu) = (u^-\cdot \tilde{A}^-_\mu) - 2\partial_\mu f^{ai}u^-_a u^-_i$.
We substitute \p{sol-eq1} and \p{sol-eq4} into \p{Lmckin}
 and, integrating by parts with respect to $D^{++}$, obtain
\beq
 {\cal L}_{kin}&=&
 -{1\over 2}\int du
 A^{a-}_\mu \partial^\mu F^+_a \nonumber \\
&=& \int du
 \left\{
 {1 \over 2}D^{++}(u^-\cdot A^-_\mu)\partial^\mu f^{ai}u^+_{(a}u^-_{i)} 
 \right. \nonumber \\
&&
 \left. -\left( \partial_\mu f^{ai}u^+_a u_i^-
 + \frac{1}{1+V_0}\partial_\mu f^{ai}\,V_{ai}\right)
 \left(\partial^\mu f^{bj}u^-_bu^+_j + \partial^\mu v\right)\right\}\,. \label{111}
\eeq
At this step, eq. \p{proj2} 
 must be taken into account. We make use of it in the first term
on the r.h.s. of \p{111}, then perform the harmonic integral,
 integrate a few times by parts
 and use eqs. \p{vanv} and \p{eq1t4}.
Finally, we obtain the kinetic sigma model term just in the
 form \p{last} with
\begin{eqnarray}
g_{ai,bj}=(1+V_0)\epsilon_{ab}\epsilon_{ij}
 +V_{ai}\epsilon_{bj}+V_{bj}\epsilon_{ai}+\frac{2}{1+V_0}
 V_{ai}V_{bj}\,.
 \label{inv-metric}
\end{eqnarray}
The same metric has been earlier derived from the HSS approach in
 \cite{hss3,hss2}.
There,
 the non-Lagrangian approach was used,
 with the inverse metric as the basic outcome:
\begin{eqnarray}
g^{ai,bj}=\frac{1}{1+V_0}(\epsilon^{ab}\epsilon^{ij}
 +V^{ai}\epsilon^{bj}+V^{bj}\epsilon^{ai}+V^2\epsilon^{ai}\epsilon^{bj})
 \label{mc-metric}
\end{eqnarray}
where $V^2=V^{ai}V_{ai}$.
The Lagrangian approach used above is simpler and more direct.
It can be easily employed to find the explicit form of the scalar potential term in \p{finact}
 for the considered multi-center metrics.

To this end, we should still solve eqs. \p{13a} and \p{13b}
 which in this particular case have the following form
\beq
&&{\cal D}^{++}M_{a}^- + im\,u^+_a = 0\,, \label{20b}\\
&&{\cal D}^{++}N_{a}^- - im\,u^+_a = 0\,. \label{20c}
\eeq
Thus eqs. \p{20b} and \p{20c} are reduced to
 the single equation
\beq
&& D^{++}M_{a}^- - u^+_a\,(u^{+}\cdot M^{-})\,L + im\,u^+_a = 0\,.\label{21b}
\eeq
Introducing
\beq
\tilde{M}^{a-} = M^{a-} + im\,u^{-a} \label{22b}
\eeq
and projecting \p{21b} on the harmonics $u^+_a$ and
 $u^-_a$, respectively, we obtain
\beq
&& D^{++} (u^+\cdot \tilde{M}^-) = 0 \,, \label{23a} \\
&& D^{++} (u^-\cdot \tilde{M}^-) - (u^+\cdot \tilde{M}^-)\left( 1 - L\right)
 - im L = 0\,. \label{23b}
\eeq
It follows from \p{23a} that $(u^+\cdot  \tilde{M}^-)$ does not depend on harmonics,
\beq
(u^+ \cdot \tilde{M}^-) = A(x)\,. \label{24a}
\eeq
Substituting this into \p{23b} and integrating the l.h.s. of 
the latter over harmonics, we obtain
\beq
A(x) = im\, \frac{V_0}{1 + V_0}\, \quad \Rightarrow \quad (u^+ M^-)
 = A -im = -im \,\frac{1}{1 + V_0}\,. \label{25a}
\eeq
Substituting \p{25a} into \p{61a}, we find the final form of the 
 scalar potential to be
\beq
{\cal L}_{pot}(x,u) = -m^2\,\frac{1}{1 + V_0}\,, \quad \partial^{++}{\cal L}_{pot} = 0\,. \label{26a}
\eeq

Thus, we have managed to solve the equations of motion
 \p{2a},~\p{eq4},~\p{13a} and \p{13b} and so have found the explicit form of
 the component bosonic action \p{finact} for the multi-center case. Both
 the HK metric and potential term in \p{finact} are expressed, by eqs.
 \p{inv-metric} and \p{26a}, in terms of the single object,
 multi-center potential  $V_0$ defined
 in \p{relation1}.

Let us derive the same result \p{26a} in another way.
Applying the general formula \p{6a} to the particular case \p{19a}
and taking into account that it follows from eq. \p{sol-eq1} that
$\delta f^{ai} = \varepsilon\,\epsilon^{ai} \;\Rightarrow \;k^{ai} =
\epsilon^{ai}$, we find
\beq
u^{+a} = - E^{a+}_{bi}k^{bi} = - E^{a+}_{bi}\,\epsilon^{bi}
 = {\cal D}^{++}E^{a-}_{bi}\,\epsilon^{bi}. \label{70a}
\eeq
Then we can rewrite the Lagrangian \p{61a} as
\beq
{\cal L}_{pot} = - im\,M^{a-}\Omega_{ac}E^{c\,+}_{bi} \epsilon^{bi}\,.
                \label{71a}
\eeq
Moreover, it is easy to find
\beq
M^{a-} = -im\, E^{a-}_{bi}\epsilon^{bi}\,.
\eeq
Now we substitute this into \p{71a}, use the definition \p{metric-def} and
take into account eq. \p{8a} and the fact that under the $u$-integral one
can integrate by parts. As the result we obtain
\beq
{\cal L}_{pot} = -{m^2\over 2} \,g_{ai, bk}\epsilon^{ai}\epsilon^{bk} =
-{m^2\over 2}\,  g_{ai, bk}\,k^{ai}k^{bk}\,,\label{73a}
\eeq
which is just the square of the norm of Killing vector
 $k^{ai} = \epsilon^{ai}$.
Using \p{inv-metric}, we find
\beq
g_{ai, bk}\epsilon^{ai}\epsilon^{bk} = {2\over 1 + V_0}\,,
\eeq
i.e., we come back to the expression \p{26a}.
This means that the scalar potential is determined by the norm of the
 Killing vector.
This fact was
 originally obtained in \cite{GTT1} by means of an on-shell formalism.

Finally, let us show that eqs. \p{inv-metric} and \p{26a}
 can be put in the standard multi-center form.
Introducing,
\begin{eqnarray}
 \vec{V}=-{i\over 2}(\vec{\tau})^{ai}V_{ai},~~~
 \vec{X}=\displaystyle\frac{i}{\sqrt{2}}(\vec{\tau})^{ai}f_{ai},~~~
 \varphi=\displaystyle\frac{1}{\sqrt{2}}\epsilon_{ai}f^{ai},~~~
 U=1+V_0,
 \label{relation3}
\end{eqnarray}
where $\vec{\tau}^{ai}$ are the
 Pauli matrices,\footnote{Here $\vec{\tau}^{ai}=\epsilon^{ab}\vec{\tau}_b^{~i}$
 where $\vec{\tau}_a^{~i}$ are the standard Pauli
 matrices.}
 we can write down Lagrangian \p{Lmc} in the form
\begin{equation}\label{nlsm_lag}
 {\cal L}
 = {1\over 2}\left\{U\partial_\mu\vec{X}\cdot
 \partial^\mu \vec{X} + U^{-1}{\cal D}_\mu \varphi {\cal D}^\mu
 \varphi - m^2 U^{-1}\right\},
\end{equation}
where $m$ has been changed by $m/\sqrt{2}$, and
 $\vec{X}=(X^1,X^2,X^3)$, $\varphi$ are real scalar fields,
 and ${\cal D}_\mu\varphi = \partial_\mu\varphi + \vec{V} \cdot
 \partial_\mu\vec{X}$.
The fields $\vec{V}$ and $1+V_0$, by their definition, satisfy
 the differential equations
\begin{eqnarray}
 \vec{\nabla} \times \vec{V}
 = \vec{\nabla} U,~~~ \Delta U = 0,~~~
 {\partial \over \partial \varphi}(1+V_0)=0.
 \label{eq:monopole}
\end{eqnarray}
The scalar potential is given by
\begin{eqnarray}
 V=m^2 U^{-1}. \label{potential}
\end{eqnarray}
This precisely coincides with what has been found
 in \cite{GTT1}.
In this parameterization, the $U(1)$ isometry \p{19a}
is realized as a shift of the coordinate $\varphi$
 of the HK manifold:
\begin{eqnarray}
 \delta \varphi = \sqrt{2}\,\varepsilon\,, \quad \delta \vec{X}=0\,.
\end{eqnarray}

Note that it is possible to extend the above consideration to
 the case of $4n$ dimensional HK manifolds whose metric
 have $n$ commuting translation isometries, i.e., to 
 the general case of toric HK metrics.
In this case, the ansatz for the analytic HK potential is
 $L_{toric}^{+4}=L^{+4}(u^{+i}q_{ai}^+,u)$ where $a=1,\cdots n$ \cite{hss2}.

Now we consider two particular examples.

{\noindent i) \it Taub-NUT case}

The analytic potential in the Taub-NUT (TN) case can be chosen as
 \footnote{The form of the TN and Eguchi-Hanson (EH)
 HK potentials can be found in Chapter 6.6.1
 of Ref. \cite{hss2} (see eqs. (6.72) and
 (6.73) there). The HK potential for the double Taub-NUT metric is
 obtained by a slight
 modification of these potentials.}
\beq
L^{+4}_{TN}(u^{+}\cdot q^+, u)  = {2 \over \lambda}\,(g^{++})^2
 =\frac{2}{\lambda}\left(\frac{L^{++}-c^{++}}
   {1+\sqrt{1+(L^{++}-c^{++})c^{--}}}\right)^2 \,,\quad
L^{++} = u^{+}\cdot q^+\,,\label{LTN2}
\eeq
where $c^{\pm\pm}=c^{ij}u_i^{\pm}u_j^{\pm}$,
 $c^2=\frac{1}{2}c^{ij}c_{ij}=1$ and $\lambda$ is a constant.
The corresponding potential $V_0$ is given by
\begin{equation}
V_0 = {1\over 2}\int du \,
      \frac{\partial^2 L^{+4}_{TN}}{\partial (L^{++})^2}
    = {1 \over 2 \lambda} \int du\,
      \frac{1}{\left[1 + (L^{++} - c^{++})
      c^{--}\right]^{3/2}}\,. \label{TNv}
\end{equation}
To compute the harmonic integral in \p{TNv}, we make use of
 the general formula derived in Ref.~\cite{integral}:
\beq
\int du \,\frac{1}{\left[1 + (G^{++} - c^{++})c^{--}/c^2 \right]^{3/2}} =
\frac{\sqrt{c^{ik}c_{ik}}}{\sqrt{G^{ik}G_{ik}}}\,,\label{gen}
\eeq
where
\beq
G^{++} = G^{ik}u^+_iu^+_k \,.
\eeq
Using this general formula and choosing the particular
 $SU(2)_R$ frame, $c^{11} = c^{22} = 0, c^{12} =
 i$, it is easy to find
\beq
U_{TN} = 1 +V_0 = 1 +
{1 \over \sqrt{2}\, \lambda}\,
 \frac{1}{\sqrt{L^{ik}L_{ik}}}\,.\label{eq:TNv}
\eeq
We can rewrite \p{eq:TNv} in terms of the multi-center coordinate $\vec{X}$.
Using the relations
\begin{eqnarray}
& L^{++}=u^{+a}F_a^+=-f^{ai}u_a^+u_i^+ = -f^{(ai)}u_{(a}^+u_{i)}^+,&
   \label{formula1} \\
& \vec{X}\cdot\vec{X}=f_{(ai)}f^{(ai)}, \nonumber&
\end{eqnarray}
we find
\begin{eqnarray}
 U_{TN} = 1 +
{1\over \sqrt{2}\, \lambda}\,
 \frac{1}{\vert\vec{X}\vert}\,. \label{fTN}
\end{eqnarray}
This form of the one-center TN potential corresponds to the center located at $\vec{X}=0$.
In the following example, 
 for later convenience, we choose another position of the center.

{\noindent ii) \it Double Taub-NUT case}\\
Now we consider more general double Taub-NUT (DTN) case.
The relevant analytic HK potential reads
\beq
L^{+4}_{DTN}(u^+\cdot q^+, u) &=&
 2\left( \frac{L^{++}}{1 + \sqrt{1 + L^{++}\eta^{--}}}\right)^2
 + {2\over \gamma}\,
 \left( \frac{L^{++}}{1 + \sqrt{1 - L^{++}\eta^{--}}}\right)^2 \nonumber \\
 && - (1-a) (L^{++})^2 \,,\label{EHv}
\eeq
where $a$ and $\gamma$ are some constants and
$\eta^{\pm\pm}=\eta^{ik}u_i^\pm u_i^\pm$. If $\eta^{ik} = 0$ we
return to the TN case. For $\eta^{ik} \neq 0$, one can always choose
 $\eta^2 = {1 \over 2} \eta^{ij}\eta_{ij}=1\,$ by the appropriate rescaling of $q^{a+}$.
As we will see, $\eta^{ij}$ specifies the location of the centers.
The potential $V_0$ in the present case reads
\beq
V_0 = {1\over 2} \int du
       \left( \frac{1}{\left[1 + L^{++}\eta^{--}\right]^{3/2}} +
       {1\over \gamma}\,\frac{1}{\left[1
        - L^{++} \eta^{--}\right]^{3/2}}\right) -(1-a) \,,
      \label{EHd}
\eeq
whence
\beq
U_{DTN}  = 1 + V_0 = a+
      {1\over \sqrt{2}}
     \frac{1}{\sqrt{(L^{ik} + \eta^{ik})(L_{ik} + \eta_{ik})}} +
      {1\over \sqrt{2}\,\gamma}\,
     \frac{1}{\sqrt{(L^{ik}-\eta^{ik})(L_{ik}-\eta_{ik})}}\,.
       \label{eq:UEH}
\eeq
Introducing,
\begin{eqnarray}
\vec{\xi}={i \over \sqrt{2}}\vec{\tau}^{ij}\eta_{ij},
\end{eqnarray}
and using \p{formula1}, we can rewrite \p{eq:UEH} as
\begin{eqnarray}
U_{DTN}  = a +
 {1\over \sqrt{2}}\,
 \frac{1}{\vert\vec{X} + \vec{\xi}\vert} +
 {1\over \sqrt{2}\,\gamma}\,
 \frac{1}{\vert\vec{X} - \vec{\xi}\vert}\,.
 \label{fEH}
\end{eqnarray}
For $a\neq 0$ this is a two-center ALE potential, with the constant vector $\vec{\xi}$ specifying
the position of both centers (they are collinear to each other).
For $a=0$, the potential $U_{DTN}$ becomes the
 general EH potential with non-equal ``masses''.
Like the DTN potential itself, its EH limiting case possesses only $U(1)\times U(1)$
isometry in contrast to the $SU(2)\times U(1)$ isometry of the standard EH potential
which is recovered under the choice $\gamma=1$.

\section{Structure of SUSY vacua and the BPS equation}
\label{sec:BPS}
{}From the form of the scalar potential \p{26a} we can find the
 SUSY vacuum condition which is similar to that in the ${\cal N}=1$ case.
The condition of SUSY vacuum is the vanishing of the scalar potential
\begin{eqnarray}
 0=V(f^{ai})={m^2 \over 1+V_0}=g_{ai,bj}k^{ai}k^{bj}\,.
\end{eqnarray}
We find that in our parameterization of the multi-center case the SUSY vacuum
exists, provided there is a point where the potential $V_0$
 goes to infinity,
\begin{eqnarray}
V_0 = {1 \over 2} \int du
 {\partial^2 L^{+4} \over \partial (L^{++})^2}
 \rightarrow \infty. \label{cond}
\end{eqnarray}
The condition \p{cond} is the ${\cal N}=2$ multi-center counterpart of the
general SUSY vacuum conditions  \p{eq:stationary_superpotential} and \p{eq:singularity_kahlermetric}
of the ${\cal N}=1$ theory. We expect that this condition imposes strong restrictions on the original HK potential
 $L^{+4}$, though for the time being we do not know them in 
 full generality.
Now we apply eq. \p{cond} to the previous examples.

{\noindent i) \it TN case}

In the TN case, it follows from \p{fTN} that the condition \p{cond} can be realized only
for the vacuum expectation value $\vec{X}=0$.
Thus the theory has only one SUSY vacuum. As a consequence, no domain wall solution
can be found in this case since the existence of the domain wall solutions requires that the theory has
at least two vacua.

{\noindent ii) \it DTN case}

In the DTN case, the theory has two discrete vacua
 which are realized at vacuum expectation values
 $\vec{X}=-\vec{\xi}$ and $\vec{X}=\vec{\xi}$.
Indeed, in this case there exists the domain wall solution as we shall see
soon.

In order to find the behaviour of the potential in the DTN case, we
 take $\vec{\xi}=(0,0,\xi)$ and
introduce the spherical coordinates such as
\begin{eqnarray}
X^1=r\sin\Theta\cos\Psi,~~X^2=r\sin\Theta\sin\Psi,
 ~~~X^3=\sqrt{r^2+\xi^2}\cos\Theta,
 ~~~\varphi=\Phi+\Psi. \label{CoordNew}
\end{eqnarray}
In this parameterization, the geometrical meaning of the target manifold
 becomes clear: the fields $r$ and $\Theta$ are the coordinates of
 base manifold $S^2$ and $\Phi$ and $\Psi$ form a fiber over this base
 manifold. The DTN potential  \p{fEH} takes the following form
 in the coordinates \p{CoordNew}:
\begin{eqnarray}
U_{DTN}=a+{1 \over \sqrt{2}}
 \left({1 \over |\sqrt{r^2+\xi^2}+\xi\cos\Theta|}
      +{1 \over \gamma}{1 \over |\sqrt{r^2+\xi^2}-\xi\cos\Theta|}\right).
 \label{spherical}
\end{eqnarray}
Note that the potential \p{spherical} depends only on the real coordinates
 $r$ and $\Theta$, and not on $\Phi$ and $\Psi$, which reflects the presence of $U(1)\times U(1)$
 isometry. So the vacuum configurations ``live'' on the submanifold $S^2$ of the full target space.

\begin{figure}[t]
\begin{center}
\leavevmode
\begin{eqnarray*}
\begin{array}{ccc}
  \epsfxsize=7.5cm
  \epsfbox{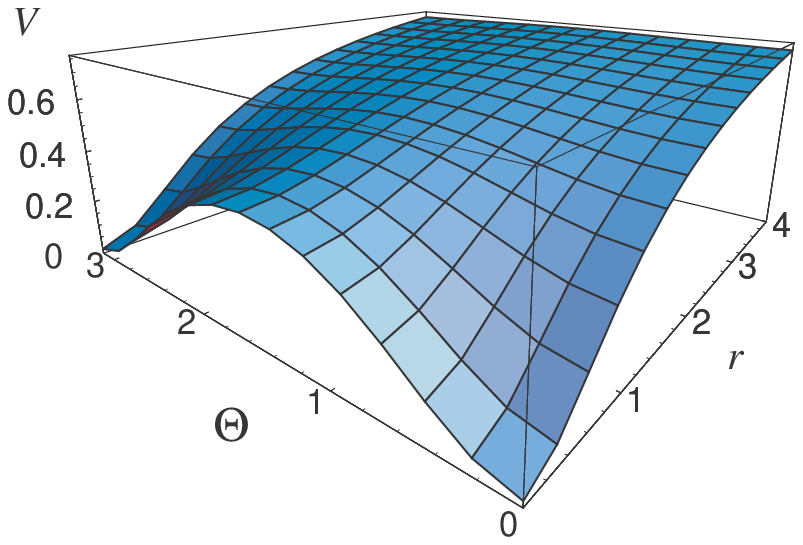} &
  \hspace{5mm}       &
  \epsfxsize=12.5cm
  \epsfbox{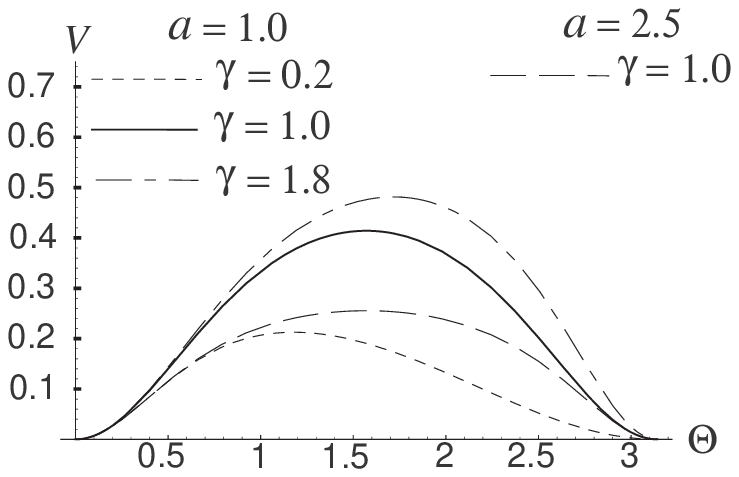} \\
\mbox{\footnotesize (a) 3D Plot of the potential.}&
 & \hspace{-4.5cm}
\mbox{\footnotesize (b) Plot of the potential along $\Theta$ axis.}
\end{array}
\end{eqnarray*}
\caption{Plots of the potential for $\xi=1.0$ and $m=1.0$.}
\end{center}
\end{figure}
\label{plot1}
The behaviour of the scalar potential in the spherical reparameterization
 for different ranges of the involved parameters is shown in Fig. 1.
Fig. 1-(a)
 shows the 3D plot of the scalar
 potential.
It is seen that there are two discrete vacua at
 $(r,\Theta)=(0,0)$ and $(0,\pi)$.
Fig. 1-(b)
 shows the plot of the potential along $\Theta$ axis
 for some values of $a$ and $\gamma$.
For $\gamma=1$ the shape of the potential is symmetric with respect
 to $\Theta=\pi/2$ axis.
As the value of $\gamma$ deviates  from $\gamma=1$, the scalar
 potential becomes asymmetric.
If $\gamma$ goes to infinity, the vacuum is realized only at
 the single point $(r,\Theta)=(0,0)$ since the third term in eq. \p{fEH}
 disappears and the potential \p{fEH} reduces to the one of the TN case
 (with an arbitrary constant $a$).
As $a$ increases, the behaviour becomes gentle, i.e., 
 the value of the potential
 at any value of $\Theta$
 excepting vacua decreases.

The vacuum expectation values, for instance, $\vec{X}=0$ in the TN case,
 can be easily cast in the
 HSS language using \p{formula1}; these amount to $L^{++}=0$ in the TN case
 and to $L^{++}=-\eta^{++}$ and $L^{++}=\eta^{++}$ in the DTN case.
For these special values of $L^{++}$, the analytic
 potentials \p{LTN2} and \p{EHv}, equally as their second derivatives
 entering \p{TNv} and \p{EHd}, acquire singularities at some points of the harmonic
 sphere $S^2 \sim \{u^+_i, u^-_k\}$. As the result, the $u$-integral \p{cond} which specifies
 the sufficient condition for vacuum to be the SUSY one becomes divergent.
E.g., in the TN case, substituting $L^{++}=0~(L^{ij}=0)$
 into \p{cond}, one obtains harmonic integral
\begin{eqnarray}
 V_0{\Big |}_{L^{ij}=0}=\frac{1}{2}
 \int du {1 \over [1-c^{++}c^{--}]^{3/2}} \label{suf}
\end{eqnarray}
which is divergent.
The same divergent integrals are  obtained in the DTN case for two values of
 $L^{++}$, namely for
 $L^{++}=-\eta^{++}$ and $L^{++}=\eta^{++}$.
The existence of two discrete vacua in the latter case guarantees
 the existence of the domain wall solution.

In the following, we consider the general BPS equation and
 apply it to the DTN case.
If we assume that there is a non-trivial configuration along
the spatial  $y$ direction and this configuration is static,
the energy density can be written as
\begin{eqnarray}
{\cal E}&=&\displaystyle\frac{1}{2}U\partial_2 \vec{X}\cdot
 \partial_2 \vec{X} + \frac{1}{2}U^{-1}{\cal D}_2 \varphi {\cal D}_2
 \varphi + \frac{1}{2}m^2 U^{-1} \nonumber \\
&=&{1 \over 2}U(\partial_2 \vec{X}- m U^{-1}\vec{n})\cdot
 (\partial_2 \vec{X} - m U^{-1}\vec{n})+\partial_2\vec{X}\cdot\vec{n}
 + \frac{1}{2}U^{-1}{\cal D}_2 \varphi {\cal D}_2 \varphi \nonumber \\
&\ge& \partial_2\vec{X}\cdot\vec{n},
\end{eqnarray}
where $\vec{n}$ is a unit vector.
BPS equation is easily read off as
\begin{eqnarray}
 &\partial_2 \vec{X} - m U^{-1}\vec{n}=0,& \label{BPSeq1}\\
 &{\cal D}_2 \varphi=0.& \label{BPSeq2}
\end{eqnarray}
Taking $\vec{n}=(0,0,1)$ and using $\vec{n} \cdot \vec{V}=0$
 \cite{papa}, BPS equation is simplified to
\begin{eqnarray}
&&\frac{\partial \varphi}{\partial y}=0,
  ~~~
  {\partial X^1 \over \partial y}={\partial X^2 \over \partial y}=0,
  \label{BPS-eq1}\\
&&\displaystyle\frac{\partial X^3}{\partial y} = m U^{-1}.
\label{BPS-eq2}
\end{eqnarray}
Eq. \p{BPS-eq1} can be easily solved by
\begin{eqnarray}
 \varphi={\rm const},
 ~~X^1={\rm const},~~X^2={\rm const}. \label{sol-BPS1}
\end{eqnarray}
Without loss of generality, these constants can be put equal to zero, 
 i.e., $\varphi=X^1=X^2=0$.
Using these solutions and substituting \p{fEH} into \p{BPS-eq2},
 we bring eq. \p{BPS-eq2} to the form
\begin{eqnarray}
{\partial X \over \partial y} =
 m {\sqrt{2}\gamma(\xi^2-X^2) \over \sqrt{2}a\gamma(\xi^2-X^2)+\gamma(\xi-X)+\xi+X},
 ~~~~~X\equiv X^3.
\end{eqnarray}
This equation can be easily solved.
Fig. 2 shows the profiles of the domain wall solutions.
Fig. 2-(a)
 shows the profiles for some values of
 $\gamma$ with $a=1$.
For $\gamma=1$, the metric becomes the DTN metric with equal masses
 and the scalar configuration is symmetric at the center of the wall for $y=0$.
In particular, for $\gamma=1$ and $a=0$ (the standard EH
 case), analytic solution is obtained as
\begin{eqnarray}
  X = \xi\tanh\left({\sqrt{2} m \over 2 \xi}(y+y_0) \right),
\end{eqnarray}
where $y_0$ is an integration constant which specifies the position of the
 wall.
As $\gamma$ deviates from the value $\gamma=1$,
 the scalar configuration becomes asymmetric.
For fixed $\gamma$, the behaviour of the solution is shown
 in Fig. 2-(b).
As was mentioned, when $\gamma\rightarrow\infty$,
 the metric approaches the TN metric and therefore
 the domain wall solution does not exist in this limit.
\begin{figure}[t]
\begin{center}
\leavevmode
\vspace{-2cm}
\begin{eqnarray*}
\begin{array}{ccc}
  \epsfxsize=7.5cm
  \epsfbox{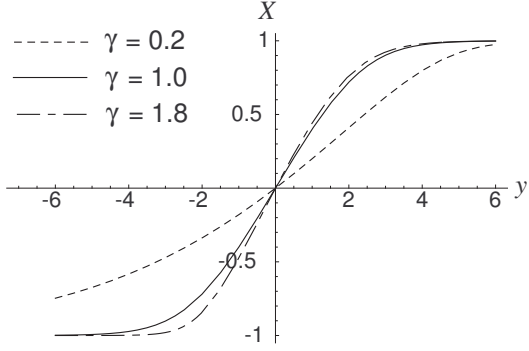} &
  \hspace{5mm}       &
  \epsfxsize=8.5cm
  \epsfbox{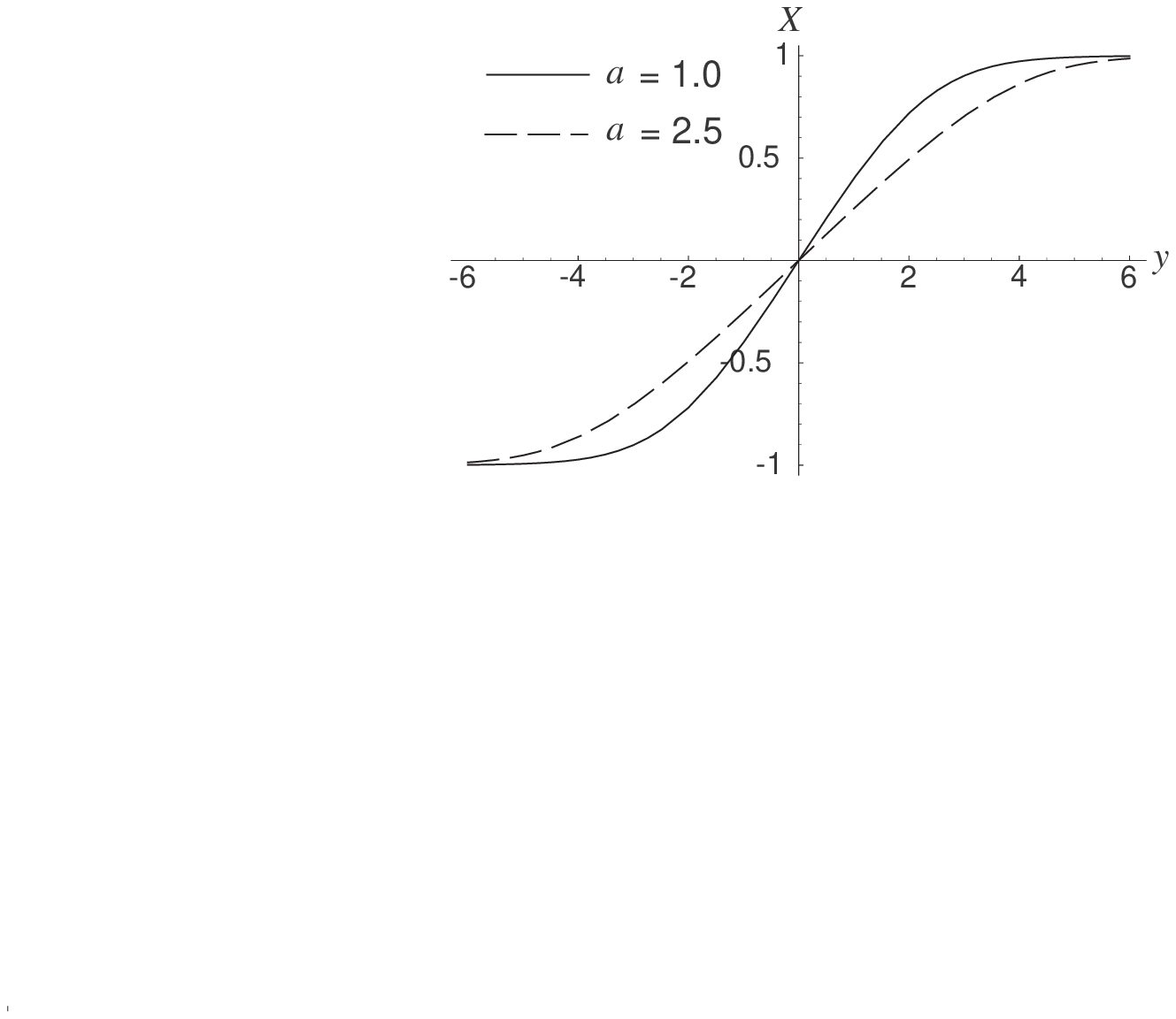} \\
\mbox{\footnotesize (a)Wall solutions for some values of $\gamma$ with $a=1$.} &
 &
\mbox{\footnotesize (b)Wall solutions for some values of $a$ with $\gamma=1$.}
\end{array}
\end{eqnarray*}
\caption{Behaviour of the domain wall solution. The wall is positioned at $y=0$.}
\end{center}
\end{figure}

\section{Summary and concluding remarks}
\label{sec:concl}
In this paper we have studied ${\cal N}=2$ massive nonlinear sigma
 model starting
 from the action in the off-shell HSS formulation which manifests the full
 ${\cal N}=2$ SUSY. The scalar potential was obtained by assigning to $q^+$ a
 dependence on the central charge coordinate $x_5$, such
 that $\partial/\partial x_5$ is identified with the
 Killing vector of the isometry.
The component bosonic action was obtained based on
 the one-to-one correspondence between the Lagrangian
 and non-Lagrangian approaches to the HK geometry.
As was shown in \cite{hss2}, the kinetic term at the component level
 in the general nonlinear sigma model
 is composed of the vielbeins and has a form which is independent
 of the harmonic variables.
On the other hand, the scalar potential in the general case
 of one isometry still involves an integration over harmonics.
Its more preferable form, which does not contain the harmonic integral,
 is derived in the Appendix B.
It is shown there that the integration over harmonic
 variables can be explicitly performed under the particular
 condition which is satisfied, e.g., in the multi-center case.

Massive nonlinear sigma models with multi-center metrics were examined.
In the generic HK case, solving the kinematical part of
 the equations of motion is very difficult problem.
However, in the multi-center case, the situation is much
 simpler.
We solved the kinematical part of the equations of motion
 and obtained the physical component action where integration over
 harmonic variables was performed to the end.
It was shown that both the target metric and the scalar potential
 can be expressed in terms of the single analytic HK potential.
The scalar potential was found to be fully specified by the norm
 of the Killing vector,
which is in agreement with the earlier derivation of Ref. \cite{GTT1}.
Given the explicit form of the scalar potential, we discussed
 the SUSY vacuum condition.
The SUSY vacuum condition was related to the analytic HK potential.
This result is the ${\cal N}=2$ extension of the similar condition in
 ${\cal N}=1$ theory which involves the K{\"a}hler potential
 and the superpotential. We derived BPS equation in the general
 multi-center case and
 BPS domain wall solution was obtained for the DTN case.

We would like to conclude
  by mentioning some directions for future work.

It would be interesting to extend our discussion of the
 multi-center four-dimensional case
 to the general case of toric HK manifolds, i.e., $4n$ dimensional HK
 manifolds with $n$ commuting triholomorphic isometries.

In this paper, we studied the nonlinear sigma model without explicit use of the quotient construction,
starting from the given HK potential $L^{+4}(q, u)$
(though both the TN and DTN HK potentials can actually be deduced from the appropriate
HSS version of the quotient construction, see \cite{EHhss,IV,CIV}).
It is interesting to discuss the massive nonlinear sigma model with isometries like
the models we considered here within the quotient construction. We expect essential simplifications
 in this regard. 
In particular, the quotient method could make easier coupling of the
massive nonlinear sigma model to the supergravity,
 which is also an interesting task.
In this case, instead of HK sigma models we
 would have sigma models with quaternion-K\"ahler (QK) target
 manifolds \cite{BW}.
The HSS formulation of the ${\cal N} = 2, d=4$ supergravity and massless QK sigma models in the supergravity background
was worked out in Refs. \cite{hssgrav,BGIO,IV}. It would be tempting to extend it to the case of
massive QK sigma models.
\vspace{1.0cm}

\noindent{\Large \bf Acknowledgements}

\noindent The collaboration with M. Naganuma, M. Nitta and N. Sakai motivated M.A. to start this work.
The work of E.I. was supported in part by grants DFG No.436 RUS 113/669, RFBR-DFG 02-02-04002, INTAS
00-00254, RFBR 03-02-17440 and grants of the Heisenberg-Landau and Votruba-Blokhintsev programs.
He thanks the Institute of Physics in Prague for warm hospitality at the initial stage of this study.

\renewcommand{\thesubsection}{\thesection.\arabic{subsection}}

\appendix

\section{Relation between the Killing vectors in harmonic and ordinary
 spaces}
In this appendix, we present the expression of the Killing
vector $k^{ai}$ in terms of the central basis Killing potential.

In order to obtain this expression, we use the formulas for
 vielbeins, which were derived in the non-Lagrangian
 geometric approach \cite{hss3}.
Firstly, we multiply both sides of the relation in \p{6a},
\begin{eqnarray}
\lambda^{a+} =  -E_{ck}^{a+}k^{ck}=
 \frac{1}{2}\Omega^{ab}\partial_{b+}f^{ci}\partial_{ci}
 \Lambda^{kl}u_{k}^+u_l^+\,,
\end{eqnarray}
by $E^{-dk}_{a}$ which is the inverse to the vielbein $E_{dk}^{a+}$.
Then we use the representation \cite{hss3}
\begin{eqnarray}
E^{-ci}_b = -\partial_{b+}f^{ci} + E^{-d-}_b\partial_{d-}f^{ci}\,,
\end{eqnarray}
where $\partial_{d-} = \partial/\partial F^{d-}$ is the derivative
with respect to the non-analytic coordinate $F^{d-}$ defined in \p{F-F+}, and
the precise form of $E^{-d-}_b$ is of no interest for our purpose.
Due to the analyticity of $\Lambda^{++}$, i.e. the property $\partial_{d-}\Lambda^{++} =0$,
the result of multiplication of \p{6a} by $E^{-dk}_a$ can be written as
\beq
E^{-dk}_a E^{a+}_{bi}k^{bi} = {1\over 2}\,\Omega^{ab}E^{-dk}_a E_b^{-ci} \partial_{ci}\Lambda^{ln}\,u^+_lu^+_n\,.
\label{27a}
\eeq
Next we make use of eq. \p{8a} and the relations
\beq
&&E^{-dk}_aE^{a+}_{bi} + E^{+dk}_a E^{a-}_{bi}
 = \delta^d_b\,\delta^k_i\,,\label{28a} \\
&&E^{+dk}_a = -{\cal D}^{++}E^{-dk}_a\,.\label{29a}
\eeq
After that, integrating both sides
 of \p{27a} over harmonics with making use of
 \p{28a}, \p{29a} and \p{8a} and taking into account
 that the $Sp(n)$ connections drop out altogether because all
 $Sp(n)$ indices are contracted, we obtain
\beq
k^{ai} = \int du \, \Omega^{cd}(E^{- ai}_c E^{- bj}_d )\,
 \partial_{bj}\Lambda^{kl}\,u^+_ku^+_l\,.\label{30a}
\eeq

Using the definition of the triplet of complex structures
\begin{eqnarray}
 I^{\pm\pm[ai,bj]}=I_{kl}^{[ai,bj]}u^{\pm k}u^{\pm l}
 =E_c^{\pm ai}E_d^{\pm bj}\Omega^{cd} \label{complex}
\end{eqnarray}
 and performing
 the harmonic integration
 in \p{30a}, we obtain the final relation
\beq
k^{ai} =
 -{1\over 3}\,I^{[ai, bj]}_{(kl)}\,\partial_{bj}
 \Lambda^{(kl)}(f)\,. \label{31a}
\eeq

\section{The scalar potential with arbitrary $U(1)$ isometry}
In this appendix, we solve the kinematical equations of motion
 \p{13a} and \p{13b}.
First of all, in order to solve them, we prove the following relation
\beq
\partial_{a+}(F^+_b\lambda^{b+})=\partial_{a+}D^{++}({\cal F}^-_b\lambda^{b+}) =
{\cal D}^{++} \partial_{a+}({\cal F}^-_b\lambda^{b+})
 + (\partial_{a-}{\cal F}^-_b)\lambda^{b+} \label{16a}
\eeq
where ${\cal F}^-_a$ was defined in eq. \p{15a}.
The first equality can be proved
by using the definition \p{36a} and \p{15a}.
In order to prove the second equality, one should firstly take
 into account that
 ${\cal F}_b^-\lambda^{b+}$ is not analytic since ${\cal F}_b^-$ is not
 analytic, ${\cal F}_b^-= {\cal F}_b^-(F^+,F^-,u)$.
We shall also need the property that the harmonic covariant derivative ${\cal D}^{++}$
 includes the derivative $\partial_{d-} = \partial/\partial F^{d-}$ when acting on non-analytic objects
 \cite{hss3,hss2}
\begin{eqnarray}
{\cal D}^{++}G_a(F^+,F^-)
 = D^{++}G_a + F^{d+}\partial_{d-}G_a
   -{1 \over 2}\partial_{a+}\partial_{b+}L^{+4}G^b\,. \label{cov1}
\end{eqnarray}
Using \p{cov1}, as well as the relations $\partial_{a-}F^{b+}=0$ and
 $\partial_{a+}F^{b+}=\delta_{a}^{~b}$,
 we can prove the second equality.

Introducing $\lambda^{a-}$ by
\beq
\lambda^{a+} = D^{++}\lambda^{a-}\;, \quad \delta F^{a-}
 = \varepsilon\,\lambda^{a-}\,,
\eeq
 and using the relation
\begin{eqnarray}
{\cal D}^{++}\partial_{[a+}{\cal F}^{-}_{b]}
 =-\Omega_{ab}+\partial_{[b-}{\cal F}^-_{a]}\,,
~~~~~~
\partial_{[a+}{\cal F}^{-}_{b]}
 \equiv {1 \over 2}(\partial_{a+}{\cal F}^-_b-\partial_{b+}{\cal F}^-_a),
\end{eqnarray}
it is straightforward to transform \p{16a} to
\beq
\partial_{a+}(F^+_b\lambda^{b+})
 = {\cal D}^{++}\left({\cal F}^-_b\partial_{a+}\lambda^{b+}
 +\lambda^{b+}\partial_{b+} {\cal F}^-_a
 + \lambda^{b-}\partial_{b-} {\cal F}^-_a \right)
 - 2 \lambda^+_a\,, \label{16aa}
\eeq
whence the solution to eqs. \p{13a} and \p{13b} is obtained as follows
\beq
M^{a-} = -N^{a-} &=& im
\left({\cal F}^-_b\partial^a_+\lambda^{b+} +\lambda^{b+}\partial_{b+} {\cal F}^{a-}
 + \lambda^{b-}\partial_{b-} {\cal F}^{a-} - 2\,E^{a-}_{ci}k^{ci} \right) \nonumber \\
&=&im\left({\cal F}^-_b\partial^a_+\lambda^{b+} + k^{ci}\left[\partial_{ci} {\cal F}^{a-}
- 2\,E^{a-}_{ci}\right] \right)\,.
\eeq
In passing from (B.5) to the first line in (B.6) we used eqs. (A.1) and (A.5).
Now we can substitute the solution (B.6) in \p{component2}
 and obtain the general form of the scalar potential \p{finalscalar}.

Another, perhaps more convenient representation for
 the scalar potential can be obtained as follows.
Using the $Sp(n)$ bridge $M^a_b\,, M^{ab}M_b^c = \Omega^{ac}\,$,
 one redefines
\beq
N^{-}_a = M^b_a\hat{N}^-_b\,, \;E^{-bi}_a = M^d_a\hat{E}^{-bi}_d\,,\;
 ({\cal D}^{++})^{~c}_a M^b_c = 0\,,\;
 D^{++}\hat{E}^{-bi}_d = \hat{E}^{+bi}_d\,, \;
 D^{++}\hat{E}^{+bi}_d = 0\,, \label{17aa}
\eeq
and rewrites the equation for $N_a^{-}$ as
\beq
D^{++}\hat{N}^-_a = im\,\hat{E}^{-bi}_a
 \partial_{bi}(F^+_c\lambda^{c+})\,. \label{17ab}
\eeq
The last equation is solved by
\beq
\hat{N}^-_a = im\,\int dw\, \frac{1}{(u^+\cdot w^+)}\,\hat{E}^{-bi}_a
 \partial_{bi}(F^+_c\lambda^{c+})\;.\label{17ac}
\eeq
Substituting this into \p{component2}, we obtain
 the following form of the scalar potential
\beq
{\cal L}_{pot} &=&
 -{im \over 2}
 \int du (M^{a-}-N^{a-})\partial_{a+}(F_b^+\lambda^{b+})
 \nonumber \\
 &=& m^2\,\int du dw\,
 \Omega^{ab}\hat{E}^{-di}_a \partial_{di}(F^+_c\lambda^{c+})\,
 \frac{1}{(u^+\cdot w^+)}\,
\hat{E}^{-ek}_b \partial_{ek}(F^+_f\lambda^{f+})\,, \label{17ad}
\eeq
where the first and second multipliers are evaluated at
 the harmonic arguments  $u$ and $w$, respectively.
Using the fact that the bridges drop out under the $Sp(n)$
 covariant contractions and the property
 that the vielbeins with ``hat'' are linear in the appropriate
 harmonics, one can rewrite \p{17ad} as
\beq
{\cal L}_{pot} = m^2\,I^{[ai, bk]}_{(st)}\, {\cal F}^{(st)}_{[ai, bk]} -
 \frac{m^2}{2}\, g^{ai, bk}\, {\cal F}_{ai, bk}\,, \label{17ae}
\eeq
where
\beq
&& {\cal F}_{ai, bk} = \int du dw\, \partial_{ai}(F^+_c\lambda^{c+})\,
 \frac{(u^-\cdot w^-)}{(u^+\cdot w^+)}\,
 \partial_{bk}(F^+_f\lambda^{f+})\,, \nonumber \\
&&
{\cal F}^{(st)}_{[ai, bk]} = \int du dw\,
 \partial_{ai}(F^+_c\lambda^{c+})\,\frac{u^{-(s}w^{-t)}}{(u^+\cdot w^+)}\,
 \partial_{bk}(F^+_f\lambda^{f+})\,.\label{17ag}
\eeq
However, it is unclear for us whether these harmonic integrals
can be manifestly computed in the general case. Only in the particular case, when
\beq
(F^+_a \lambda^{a+}) = -{1\over 2}(F^{a+}\partial_{a+})\Lambda^{++}
 = g \Lambda^{++}\,,\label{17af}
\eeq
with $g$ being a constant, these integrals can be computed
 in terms of $\partial_{ai}\Lambda^{(kl)}$ (using
 the fact that $\Lambda^{++}$ is bilinear in harmonics
 in the central basis), although we do not quote here the precise formulas. Note that this condition
 is trivially satisfied for the isometry \p{19a}.

%
%
%
%

\end{document}